\documentclass[aps,prd,twocolumn,superscriptaddress,notitlepage,showpacs,nofootinbib,longbibliography]{revtex4-1}

\usepackage{graphicx}
\usepackage[utf8]{inputenc} 
\usepackage{placeins}
\usepackage{amsmath}
\usepackage{amssymb}
\usepackage{float}
\usepackage{comment}
\usepackage{multirow}
\usepackage{soul}

\usepackage[usenames,dvipsnames]{color}

\def\beq{\begin{equation}}
\def\eeq{\end{equation}}
\def\bea{\begin{eqnarray}}
\def\eea{\end{eqnarray}}

\usepackage{amsbsy}
\usepackage{bm}
\usepackage{color}
\usepackage{fancyhdr}
\usepackage[pdftex]{epsfig}
\usepackage[colorlinks=true,linkcolor=blue,citecolor=red,urlcolor=magenta,filecolor=blue]{hyperref}
\usepackage{slashed}

\begin{document}

\bigskip

\vspace{2cm}

\title{Tests of the Atomki anomaly in lepton pair decays of heavy mesons}
\vskip 6ex

\author{G. L\'{o}pez Castro}
\email{glopez@fis.cinvestav.mx}
\affiliation{Departamento de F\'isica, Centro de Investigaci\'on y de Estudios Avanzados, 
Apartado Postal 14-740, 07000 M\'exico D.F., M\'exico}

\author{N\'{e}stor Quintero}
\email{nestor.quintero01@usc.edu.co}
\affiliation{Facultad de Ciencias B\'{a}sicas, Universidad Santiago de Cali, Campus Pampalinda, Calle 5 No. 62-00, C\'{o}digo Postal 76001, Santiago de Cali, Colombia}
\affiliation{Departamento de F\'{i}sica, Universidad del Tolima, C\'{o}digo Postal 730006299, Ibagu\'{e}, Colombia}

\bigskip

\begin{abstract}
The anomalies recently reported in lepton pair transitions of $^8$Be$^*$ and $^4$He nuclei may be attributed to the existence of a feebly interacting light vector boson $X17$. We study the effects of this hypothetic  particle in the semileptonic $H^* \to H e^+e^-$ decays ($H$ a $Q\bar{q}$ meson) in the framework of the HQET+VMD model. Using current bounds and the universality assumption of the $X17$ boson to quarks, we find that decays of $D^{*+}$ and $D_s^{*+}$ mesons can be  importantly enhanced relative to the dominant photon-mediated contributions. Dedicated experimental searches at current heavy meson factories may confirm the existence of this light boson or set stronger bounds of their couplings to ordinary matter.
\end{abstract}

\maketitle

\section{Introduction}

The existence of a light vector boson weakly coupled to Standard Model (SM) fermions, has been suggested as a solution to the observed discrepancy between the SM prediction and the experimental measurement of the muon $g-2$ magnetic moment anomaly (see for example~\cite{Fayet:2007ua,Pospelov:2008zw}). It may be also a good candidate as a mediator of dark and ordinary matter interactions~\cite{Fayet:2007ua,Pospelov:2008zw}. Several strategies aiming their detection in different  collider and fixed target experiments have not  found any signal so far~\cite{Essig:2013lka,wfip2020}, but  have excluded different regions in the mass and coupling strenghts of parameter space. Theoretically, different models can accomodate a light vector boson and its required interactions through dimension-four  kinetic mixing with SM neutral gauge bosons and their interactions with fermionic currents of SM or dark matter particles~\cite{Fayet:2007ua,Pospelov:2008zw,Essig:2013lka}.

The anomalies recently reported  in the invariant-mass spectrum and angular distribution of lepton pairs produced in  $^8$Be$^*$ transitions to its ground state \cite{Krasznahorkay:2015iga}, reinforces the interest in searches of light vector bosons. The observed anomalies seems to require the existence of a spin-1 boson  named $X17$ \cite{Krasznahorkay:2015iga,Feng:2016jff,Feng:2016ysn} with mass $m_X=(16.7 \pm 0.35\pm 0.50)$ MeV and a relative ratio $B(^8{\rm Be}^* \to ^8{\rm \! Be} X)/B(^8{\rm Be}^* \to ^8{\rm \! Be} \gamma) =5.8\times 10^{-6}$~\cite{Feng:2016ysn}. Couplings to standard model first-generation fermions of $\mathcal{O}(10^{-3})$ (in units of the the electron charge),  required to explain this ratio is not  discarded by  other data. More recently, the same group seems to confirm the $X17$ particle in studies of the $0^-\to 0^+$ transitions of $^4$He \cite{Krasznahorkay:2019lyl}. Several new physics extensions of the SM have been proposed in the literature with the required couplings to interpret the Atomki anomaly, including  enlarged Higgs and/or gauge sectors (see, for instance, Refs.~\cite{Feng:2016ysn,DelleRose:2017xil,DelleRose:2018eic,DelleRose:2018pgm,Seto:2020jal,Nomura:2020kcw}). Despite the excitement generated by these anomalies, one must be warned that the addition of radiative corrections to the  leading one photon exchange amplitude may be responsible \cite{Aleksejevs:2021zjw} for generating the bumps reported in the angle and mass spectrum of electron-positron pairs in $^8$Be$^*$ transitions.

The almost isosinglet nature and the small mass difference of  nuclei involved in $^8$Be$^*$ decay provides an ideal place to observe this light boson, in case it exists. Mixing of nuclear isospin states~\cite{Krasznahorkay:2019lyl, Feng:2016ysn, Feng:2020mbt} and other nuclear interference effects~\cite{Zhang:2017zap} can only partially explain the observed anomaly.  Further studies in analogous systems will be very important in order to establish or discard this light boson. In the present letter, we propose the study of $H^* \to He^+e^-$ decays, where $H(H^*)$ is a heavy $Q\bar{q}$ spin-0 (spin-1) meson. Previous related studies include: 1)  $J/\psi\to \eta_cX$  decays and associated production of $J/\psi$ mesons at BESIII and Belle II experiments, recently reported in \cite{Ban:2020uii} and, 2) a search proposal at LHCb of $D^{*0}\to D^0A'\to D^0e^+e^-$ with displaced  vertex or resonant production of the dark photon $A'$ was detailed in Ref. \cite{Ilten:2015hya}. $H^*\to He^+e^-$ decays seem to be interesting to further test the Atomki anomaly: on the one hand, the mass-splitting in heavy mesons is large enough (see Table \ref{Table:1}) to produce the $X17$ boson on-shell; on the other hand, strong decays of $H^*$ are either very suppressed of forbidden by kinematics, leaving electromagnetic decays as dominant. Furthermore, the large amount of data produced at heavy meson factories would allow to test the proposed channels in the near future. 

\begin{table*}[!t]
\centering
\renewcommand{\arraystretch}{1.2}
\renewcommand{\arrayrulewidth}{0.8pt}
\begin{tabular}{lcccc}
\hline\hline
Transition & $\ \ m_{H^*}-m_H$ (MeV) & $e_Q/m_{H^\ast}\ [{\rm GeV}^{-1}]$ & $e_q/m_{q}(0) \ [{\rm GeV}^{-1}]$ &  \ \ \ $F_{H^\ast H\gamma}(0) \ [{\rm GeV}^{-1}]$   \\
\hline
$D^{\ast +} \to D^{+} \gamma $ & \ \ \ \ \ 140.603(15) & $0.33$ & $-0.85$ & $-0.54 \pm 0.05$ \  \big[$-0.47 \pm 0.06$~\cite{PDG2020}\big] \\
$D^{\ast 0} \to D^{0} \gamma $&  \ \ \ \ \ 142.014(30) & $0.33$ & $1.70$ & $2.11 \pm 0.10$ \   \big[$< 10.8$~\cite{PDG2020}\big]\\
$D_s^{\ast +} \to D_s^{+} \gamma $ & \ \ \ \ \ 143.8(4) & $0.32$ & $-0.48$ & ($-0.17 \pm 0.03$) \ \big[$> -16.4$~\cite{PDG2020}\big] \\
$B^{\ast +} \to B^{+} \gamma $ & \ \ \ \ \ 45.37(21) & $-0.063$ & $1.70$ & $1.64 \pm 0.09$ \\
$B^{\ast 0} \to B^{0} \gamma $& \ \ \ \ \ 45.37(21) & $-0.063$ & $-0.85$ & $-0.92 \pm 0.12$ \\
$B^{\ast 0}_s \to B_s^{0}\gamma$ & \ \ \ \ \ 48.6($^{+1.8}_{-1.5}$) &$-0.062$ & $-0.48$ & ($-0.42 \pm 0.02$)  \\
\hline \hline
\end{tabular} 
\caption{\small  Mass splittings of heavy mesons and electromagnetic couplings of $H^\ast \to H\gamma$ transitions in the HQET+VMD model. Within square brackets we show experimental values when available.}
\label{Table:1}
\end{table*}

The Lagrangian describing the interaction of quark and lepton flavors $f$  with the photon and the $X$ boson is ${\cal L}_{(\gamma,X)ff}=-e\sum_f(e_fA_{\mu}+\varepsilon_{f}X_{\mu})\bar{f}\gamma^{\mu}f$, with couplings strenghts $e_f$ and $\varepsilon_f$ given in units of the electron charge $e$. The photon and $X$ boson couplings to hadrons are described each by a single vector form factor which takes into account their structure in the momentum transfer region $4m_e^2 \leq q^2\leq (m_{H^*}-m_H)^2$, with $q=p_{e^+}+p_{e^-}$.  
The form factors describing the couplings of the off-shell vector particles ($V=\gamma, X$) in $H^*(p_{H^*},\epsilon_{H^*})\to H(p_H)V(q)$ are defined from the hadronic amplitude
\begin{equation}
{\cal M}_{\mu} =ie F_{H^* HV}(q^2) \epsilon_{\mu \nu \alpha \beta}\epsilon_{H*}^{\nu}p_H^{\alpha}p_{H*}^{\beta}\ .
\end{equation}
For on-shell vector particles $V$, this Lorentz-vector amplitude must be contracted with its vector polarization $\epsilon^{\mu}_V(q)$. The case of lepton pair production is discussed in Section III.

\section{$H^*H$-Vector vertices}

The form factors $F_{H^*HV}(q)$ are evaluated in the framework of the heavy quark effective theory suplemented with vector meson dominance model (HQET+VDM) ~\cite{Colangelo:1993zq, Casalbuoni:1992dx}, which has shown to give a good description of  $H^*\to H\gamma$ decays. Since we will normalize results for our observables to this radiative decay, we use the ratio of decay rates because they are rather insensitive to the specific $q^2$-dependency of the form factor. This is due to the smallness of the $H^*-H$ mass splitting (see Table \ref{Table:1}) compared to typical hadronic scales ($\sim 1\ {\rm GeV}^2$). Also, since the contributions of heavy quarks are  $1/m_Q$ suppressed, we expect that such ratios are relatively independent of constants involved in light-quark contributions through vector meson dominance model.

For self-containess purposes, we reproduce here the term of the Lagrangian density relevant for our calculations and definitions of couplings constants \cite{Colangelo:1993zq, Casalbuoni:1992dx}. The strong interaction of heavy mesons are described by 
\begin{equation*}
{\cal L}_2(H^*HV)= i\lambda\langle {\cal H}_b\sigma^{\mu\nu}F_{\mu\nu}(\rho)_{ba}\overline{\cal H}_a\rangle\ , 
\end{equation*}
where $\langle \cdots \rangle$ denotes the trace in flavor space, $F_{\mu\nu}(\rho)=\partial_{\mu}\rho_{\nu}-\partial_{\nu}\rho_{\mu}+[\rho_{\mu}, \rho_{\nu}]$ is the field strenght tensor and $\rho^{\mu}=ig_V \widehat{\rho}^{\mu}/\sqrt{2}$ where $\widehat{\rho}^{\mu}$ the 3$\times$3 matrix of the nonet of light vector mesons. The heavy meson field ${\cal H}$ is defined in terms of the pseudoscalar ($P_a$) and vector ($P^*_{a\mu}$) mesons fields as ${\cal H}_a=\frac{1}{2}(1+\slashed{v})[P^*_{a\mu}\gamma^{\mu}-P_a\gamma_5]$, and $\overline{\cal H}_a=\gamma^0 {\cal H}_a^{\dagger}\gamma^0$. On the other hand, the coupling of light vector mesons to the vector currents are described in terms of a single constant $f_V$ in the SU(3) flavor symmetry \cite{Colangelo:1993zq, Casalbuoni:1992dx}:
\begin{equation*}
\langle 0 |\bar{q}T^i\gamma^{\mu}q|V(q,\eta)\rangle =\eta^{\mu}f_V {\rm Tr}(VT^i)\ ,
\end{equation*}
where $(T^i)_{mn}=\delta_{i m}\delta_{i n}$ and $i=1,2,3$ for $q=u,d,s$ quarks, respectively. The values of coupling constant are given below. 

The vector $H^*$ and pseudoscalar $H$ heavy mesons are composed of a $Q\bar{q}$ pair, with $Q=b, c$ and $q=u,d,s$. The hadronic matrix element of the electromagnetic current is given by \cite{Colangelo:1993zq}:
\begin{eqnarray}
 & & \left\langle H(P_H)| J_\mu^{\rm em} | H^\ast (P_{H^\ast},\epsilon_{H^\ast}) \right\rangle \nonumber \\
&=&  e\left\langle H(P_H)| e_Q \bar{Q} \gamma_\mu Q + e_q \bar{q} \gamma_\mu q | H^\ast (P_{H^\ast},\epsilon_{H^\ast}) \right\rangle, \nonumber \\
&=& e(e_Q J_{\mu}^Q + e_q J_{\mu}^q ),
\end{eqnarray}
where $e_Q(e_q)$  is the electric charge of the heavy quark (light quark) in units of the positron charge, and similarly,  $\left\langle H(P_H)| J_\mu^{\rm X} | H^\ast (P_{H^\ast},\epsilon_{H^\ast}) \right\rangle
= e(\varepsilon_Q  J_{\mu}^Q + \varepsilon_q J_{\mu}^q)$ for the $X$ boson current.

  A straightforward evaluation of the form factors in the HQET+VMD model~\cite{Colangelo:1993zq} leads to
\begin{eqnarray}\label{ffp}
F_{H^\ast H\gamma}(q^2) &=& \sqrt{\frac{m_{H^\ast}}{m_{H}}}\left[\frac{e_Q}{m_{H^\ast}} + \frac{e_q}{m_q(q^2)}  \right],  \\
F_{H^\ast HX}(q^2) &=& \sqrt{\frac{m_{H^\ast}}{m_{H}}}\left[\frac{\varepsilon_Q}{m_{H^\ast}} + \frac{\varepsilon_q}{m_q(q^2)}  \right],  \label{ffv}
\end{eqnarray}
with the effective light ``quark mass" parameter
\begin{equation}
m_q (q^2)^{-1} = - \sum_V\Big(2\sqrt{2} g_V \lambda \frac{f_V}{m_{V}^2} \Big)  \left(1- \frac{q^2}{m_V^2} \right)^{-1}\ .    
\end{equation}
The expressions for the form factors  of heavy mesons are explicitly separated in Eq. (\ref{ffp}-\ref{ffv}) into its heavy and light quark components. In the model under consideration, the couplings of heavy quarks to the the photon and $X$ boson are fixed by HQET, while the couplings to the light antiquarks are modeled by the dominance of light vector mesons \cite{Colangelo:1993zq}. For the latter,
 the sum extends over light vector-meson resonances ($V=\rho^0,\ \omega,\ \phi$) according to the light-quark content of heavy mesons ~\cite{Colangelo:1993zq}. Under the assumption of the ideal mixing for vector mesons, the couplings of light  $u$ and $d$ quarks are dominated by the exchange of  $\rho$ and $\omega$ mesons, while the coupling of the $s$ quark corresponds to the exchange of the $\phi$ meson.  

Numerical inputs for couplings constants can be found in Ref  \cite{Colangelo:1993zq} and are reproduced here for reference: $g_V=5.8$, $\lambda = -0.289 \pm 0.016\ {\rm GeV}^{-1}$ (updated from new experimental inputs \cite{PDG2020}) and $f_V$ ($m_V$) the decay constant (mass) of vector meson $V$.  Using current experimental data  for lepton-pair decays of vector mesons  $V \to e^-e^+$~\cite{PDG2020}, one gets  $(f_\rho, f_\omega, f_\phi) = (0.171, 0.155, 0.232) \ {\rm GeV}^2$, with very small uncertainties. In Table \ref{Table:1} we list values for the electromagnetic form factor predicted in the HQET+VMD model at $q^2=0$.  The quoted uncertainty is dominated by the input on the $H^*HV$ strong coupling ($\lambda$) in this model (in all the predictions from this model quoted below, all the other uncertainties are very small). A comparison with the magnitude of the measured form factor (within square brackets), obtained from the measurement of the radiative decay $D^{\ast +} \to D^{+} \gamma$  branching fraction \cite{PDG2020}, give confidence on this model.  

\begin{table*}[!t]
\centering
\renewcommand{\arraystretch}{1.2}
\renewcommand{\arrayrulewidth}{0.8pt}
\begin{tabular}{ccccc}
\hline\hline
Transition & $\varepsilon_Q/m_{H^\ast} \ [{\rm GeV}^{-1}]$ & $\varepsilon_q/m_{q}(m_X^2) \ [{\rm GeV}^{-1}]$ & $F_{H^\ast HX}(m_X^2) \ [{\rm GeV}^{-1}]$ & $ \ \ R_{X/\gamma}(H^*)$ \\
\hline
$D^{\ast +} \to D^{+} X$ & $1.84\times 10^{-3}$ & $-1.89\times 10^{-2}$ & $(-1.76\pm 0.11) \times 10^{-2}$ & \ \ $1.1 \times 10^{-3}$ \\
$D^{\ast 0} \to D^{0} X$ & $1.84\times 10^{-3}$ & $9.43\times 10^{-3}$ & $(1.17 \pm 0.05) \times 10^{-2}$ & \ \ $3.0 \times 10^{-5}$ \\
$D_s^{\ast +} \to D_s^{+} X$ & $1.75\times 10^{-3}$ & $-7.83\times 10^{-3}$ & $(-0.91 \pm 0.06) \times 10^{-2}\ \ $ & $3.1 \times 10^{-3}$ \\
$B^{\ast +} \to B^{+} X$ & $-1.39\times 10^{-3}$ & $9.43\times 10^{-3}$ & $(0.81 \pm 0.05) \times 10^{-2}$ & \ \ $1.9 \times 10^{-5}$ \\
$B^{\ast 0} \to B^{0} X$ & $-1.39\times 10^{-3}$ & $-1.88\times 10^{-2}$ & $(-2.03 \pm 0.10) \times 10^{-2}$ & \ \ $4.0 \times 10^{-4}$\\
$B_s^{\ast 0} \to B_s^{0} X$ & $-1.37\times 10^{-3}$ & $-7.83\times 10^{-3}$ & $(-0.92 \pm 0.04) \times 10^{-2}\ \ $ & $4.1 \times 10^{-4}$\\
\hline\hline
\end{tabular} 
\caption{\small The $H^*HX$ form factors evaluated at $q^2=m_X^2$ and ratio $R_{X/\gamma}$ defined in Eq. (\ref{ratioX}).}
\label{Table:2}
\end{table*}

 Let us define the following ratio of two-body decay rates:
\begin{equation}
 R_{X/\gamma}(H^*)= \frac{\Gamma(H^*\to HX)}{\Gamma(H^*\to H\gamma)}=\left|\frac{ F_{H^*HX}(m_X^2)}{F_{H^*H\gamma}(0)} \right|^2\cdot \frac{\left|\vec{p}_X\right|^3}{\left|\vec{p}_\gamma \right|^3}\ ,
\label{ratioX}
\end{equation}
where $\vec{p}_V$ is the momentum of the final state boson in the rest frame of $H^*$. 
This ratio exhibits two important differences with respect to the similar ratio defined in $^8{\rm Be}^* \to\ ^8{\rm Be}$ nuclear transitions \cite{Feng:2016jff}. First, since  $m_{H^*} \gg m_q(q^2)$ we have a suppression of the  heavy quark   relative to the light quarks contributions in Eqs. (\ref{ffp}) and (\ref{ffv}), which is stronger for bottom meson transition amplitudes. In order to be more explicit, and for the easy reference of the interested reader, in Table 2 we display the values of the two contributions that appear within square brackets in Eq. (\ref{ffv}), by assuming $q^2=m_X^2 $ for the square of the momentum transfer of the $X$-boson.  This has the advantage that the the ratio $R_{X/\gamma}(H^*)$ is more sensitive to the $Xq\bar{q}$ couplings, which are relatively well bounded from other processes \cite{ Feng:2016jff}. On the other hand, given the larger phase-space in heavy meson decays, this ratio is not suppressed by kinematics, as it happens for decay of $^8$Be nucleus.

Predictions for the $H^* \to HX$ decay fractions require an estimate of the $\varepsilon_{Q,q}$ couplings. For the couplings of the $X17$ boson to the quarks of the first generation we use: $\varepsilon_u  \simeq \pm 3.7 \times 10^{-3}$ and $\varepsilon_d \simeq \mp 7.4 \times 10^{-3}$~\cite{Fornal:2017msy}. They are obtained by combining $|\varepsilon_u+\varepsilon_d|\approx 3.7\times 10^{-3}$, obtained in Refs.~\cite{Feng:2016jff, Feng:2016ysn} from the $^8$Be$^*$ anomaly, with the null results on searches of the $\pi^0\to X\gamma$ by the NA48/2 experiment~\cite{Batley:2015lha}, which translates into the contraint $|2\varepsilon_u+\epsilon_d|\leq 8\times 10^{-4}$~\cite{Fornal:2017msy} for the $X17$ boson couplings. By assumming the NA48/2 constraint to be exactly zero, namely  the `protophobic' assumption (see however~\cite{Zhang:2020ukq}), one gets the results used in this paper. On the other hand, the limits on the coupling to electrons can be obtained $0.2\times 10^{-3} \lesssim |\varepsilon_e| \lesssim 1.4 \times 10^{-3}$  from beam dump experiments at SLAC and measurements of the $g-2$ anomalous magnetic moment of the electron according to Ref.~\cite{Fornal:2017msy}.  Our study requires the knowledge of second- and third-generation couplings, namely strange $\varepsilon_{s}$, charm $\varepsilon_{c}$, and bottom $\varepsilon_{b}$.  {\it A priori} these parameters are independent \cite{Feng:2016ysn}, and need not  be related to the first-generation couplings. Our  simplest starting assumption is  \textit{universality} of down- and up-type quark $\varepsilon_f$ couplings, thus, we will take $\varepsilon_{c}=\varepsilon_{u}$ and $\varepsilon_{b}=\varepsilon_{s}=\varepsilon_{d}$;   henceforth, our results will be obtained under this assumption \cite{Feng:2016jff,Feng:2016ysn}. Values of the $H^\ast HX$ couplings and   $R_{X/\gamma}(H^*)$ ratios for these transitions are given in Table \ref{Table:2}. The ratios are larger than the ones in the  nuclear case  mainly due to the unsuppressed phase space for $X17$ production.

\section{Lepton pair production}

The decay amplitude for lepton pair production $H^\ast(P_{H^\ast}) \to H(P_H)e^+(p_+)e^-(p_-)$ is the coherent sum of the photon and X-boson mediated amplitudes $\mathcal{M}(H^\ast \to H e^+e^-) = \mathcal{M}_\gamma +\mathcal{M}_X$, where ($V=\gamma, X$):
\begin{equation}
\mathcal{M}_V = - e^2  G_{H^\ast HV}(q^2) \ \epsilon_{\mu\nu\alpha\delta} \ell^{\mu} \epsilon^\nu_{H^\ast} P_{H}^\alpha P_{H^\ast}^\delta,
\label{amp-ee}
\end{equation}
\noindent where $\ell_{\mu}=\bar{u}(p_-)\gamma_{\mu}v(p_+)$ is the leptonic current and $G_{H^*H\gamma}(q^2)=-F_{H^*H\gamma}(q^2)/q^2$, $G_{H^*HX}(q^2)=\varepsilon_e F_{H^*HX}(q^2)/(q^2-m_X^2+im_X\Gamma_X)$. In numerical evaluations throughout  this paper we  use $\alpha_{\rm em}=\alpha(0)$, the fine structure constant, because according to Table \ref{Table:1} the maximum value of the squared photon momentum is not large ($q^2_{\rm max} = (m_{H^*}-m_H)^2$).  On the other hand, running effects between $q^2=0$ and $q^2_{\max}$ are very small compared with the present and forthcoming experimental accuracies which, in the absence of real estimates,  we will assume  to be not better than 5\% for the branching fractions.

As in Ref. \cite{Feng:2016jff,Feng:2016ysn}, we assume negligible decays of the $X17$ boson into  neutrino channels,  such that its full width is given by 
\begin{eqnarray}
\Gamma_X &\equiv& \Gamma(X \to e^+e^-) \nonumber \\ 
&=& \frac{\alpha_{\rm em}\varepsilon_e^2 m_X}{3}(1+2r_e)\sqrt{1-4r_e}\nonumber \\ 
&=&8.0 \times 10^{-8}\ {\rm MeV}
\label{nwX}
\end{eqnarray}
with $r_e=m_e^2/m_X^2$. The total width  quoted above corresponds to maximun value of $\varepsilon_e$, discussed in the previous section. Decays of a light vector boson into neutrino-antineutrino pairs that may increase  $\Gamma_X$ width are also allowed by kinematics and are included in some extensions of the SM involving enlarged Higgs and/or gauge sectors \cite{Feng:2016ysn,DelleRose:2017xil,DelleRose:2018eic,DelleRose:2018pgm,Seto:2020jal,Nomura:2020kcw}. The relevant coupling $\varepsilon_{\nu}$ can be constrained from neutrino-electron scattering in the case of the first generation like done from the TEXONO experiment~\cite{Deniz:2009mu}  yielding to $|\varepsilon_e \varepsilon_{\nu}|^{1/2} \lesssim 7 \times 10^{-5}$~\cite{Feng:2016ysn}. The addition of the $\nu\bar{\nu}$ channels will modify the total width of the $X$ boson by less that 0.1\%, and our results will remain  unchanged.

\begin{figure*}[!t]
\centering 
\includegraphics[scale=0.25]{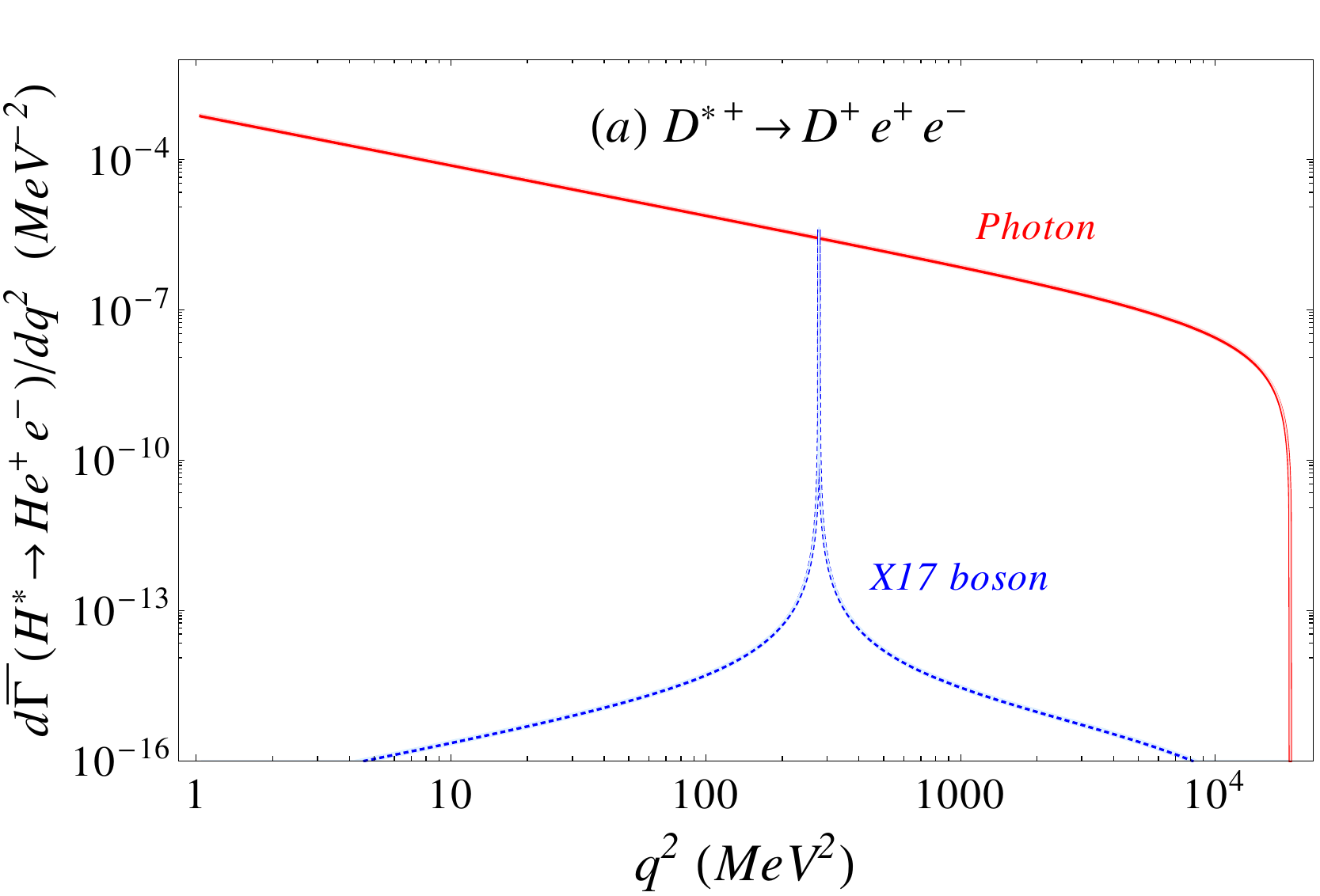} \ \
\includegraphics[scale=0.25]{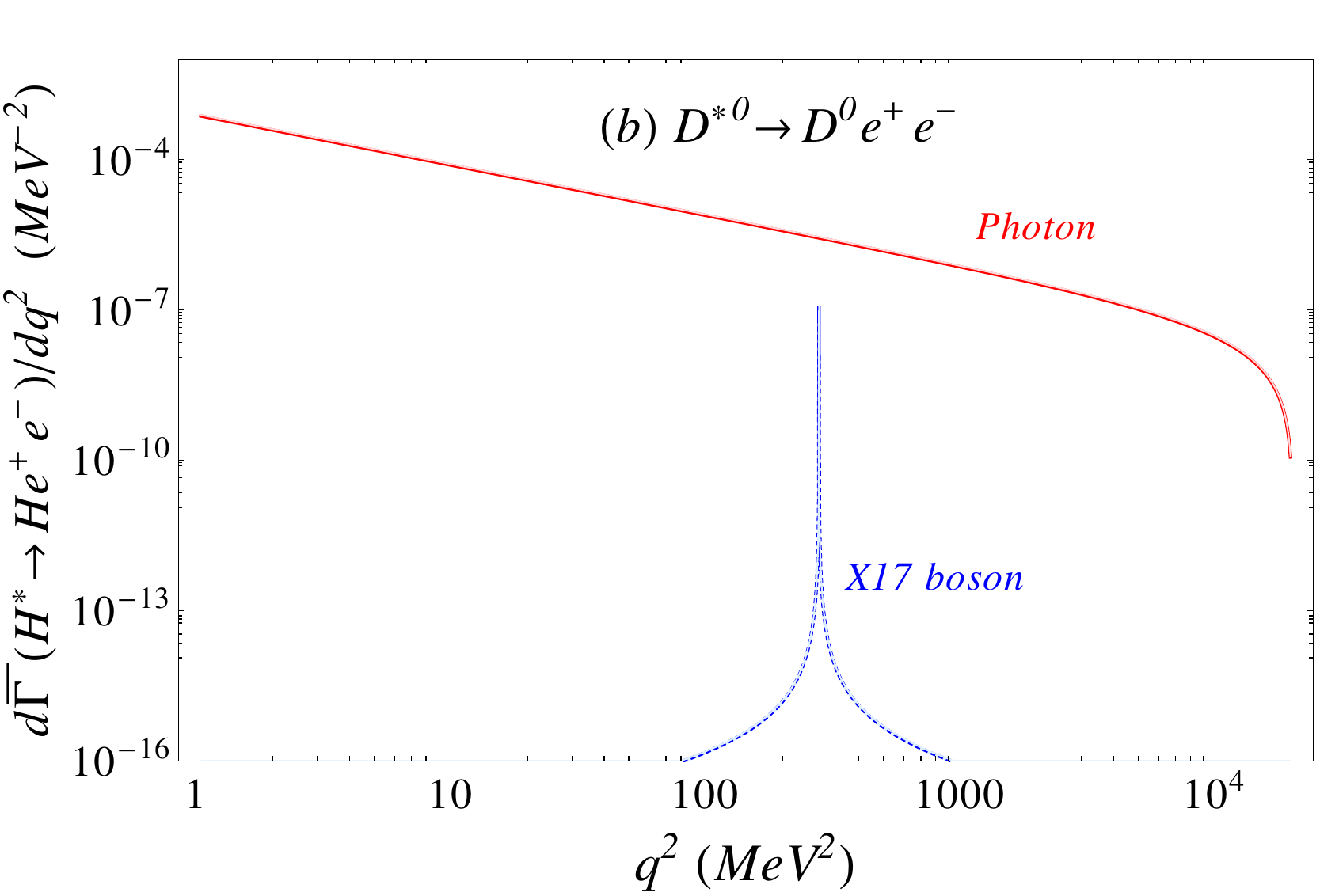} \ \ 
\includegraphics[scale=0.25]{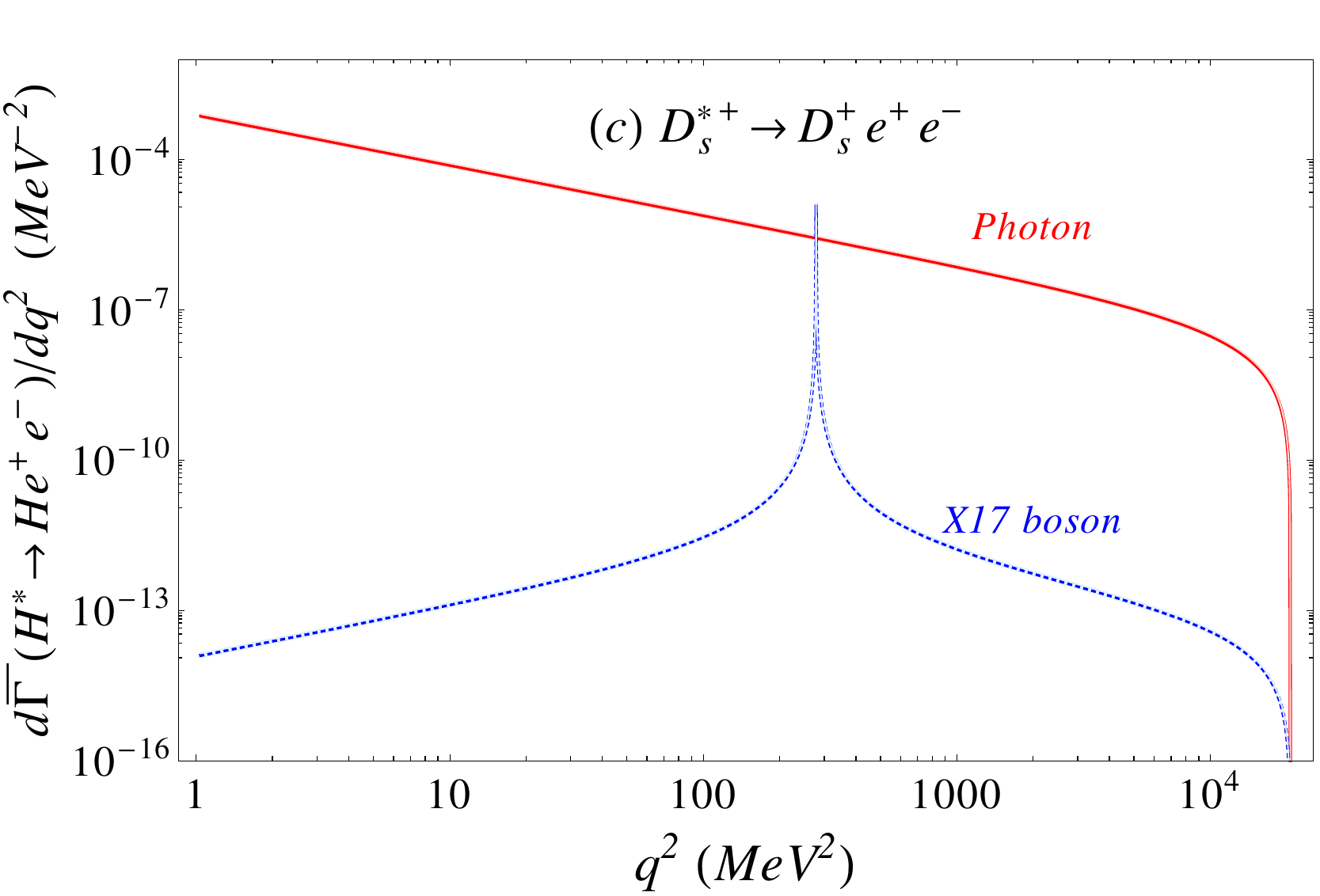} \ \
\includegraphics[scale=0.25]{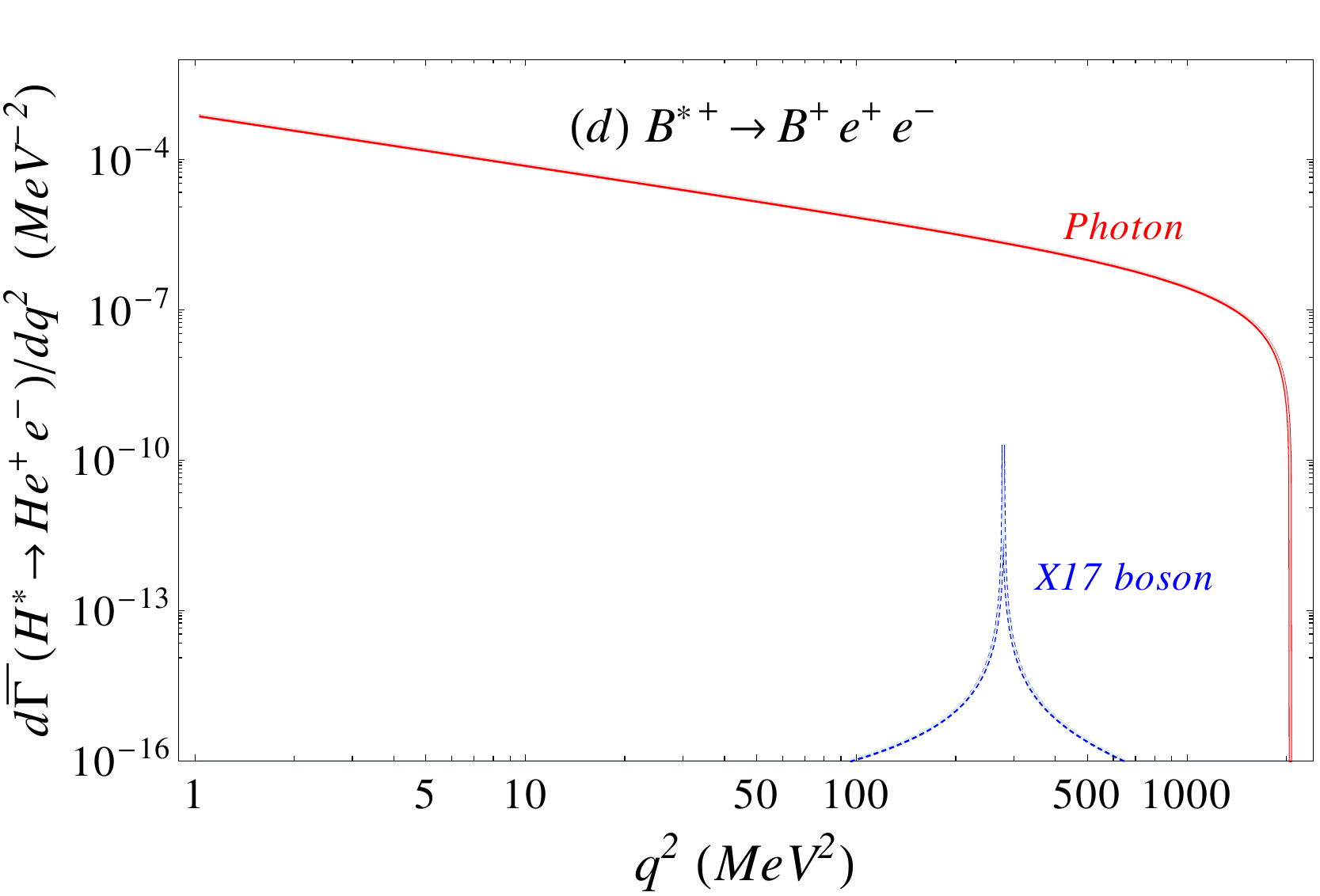} \ \
\includegraphics[scale=0.25]{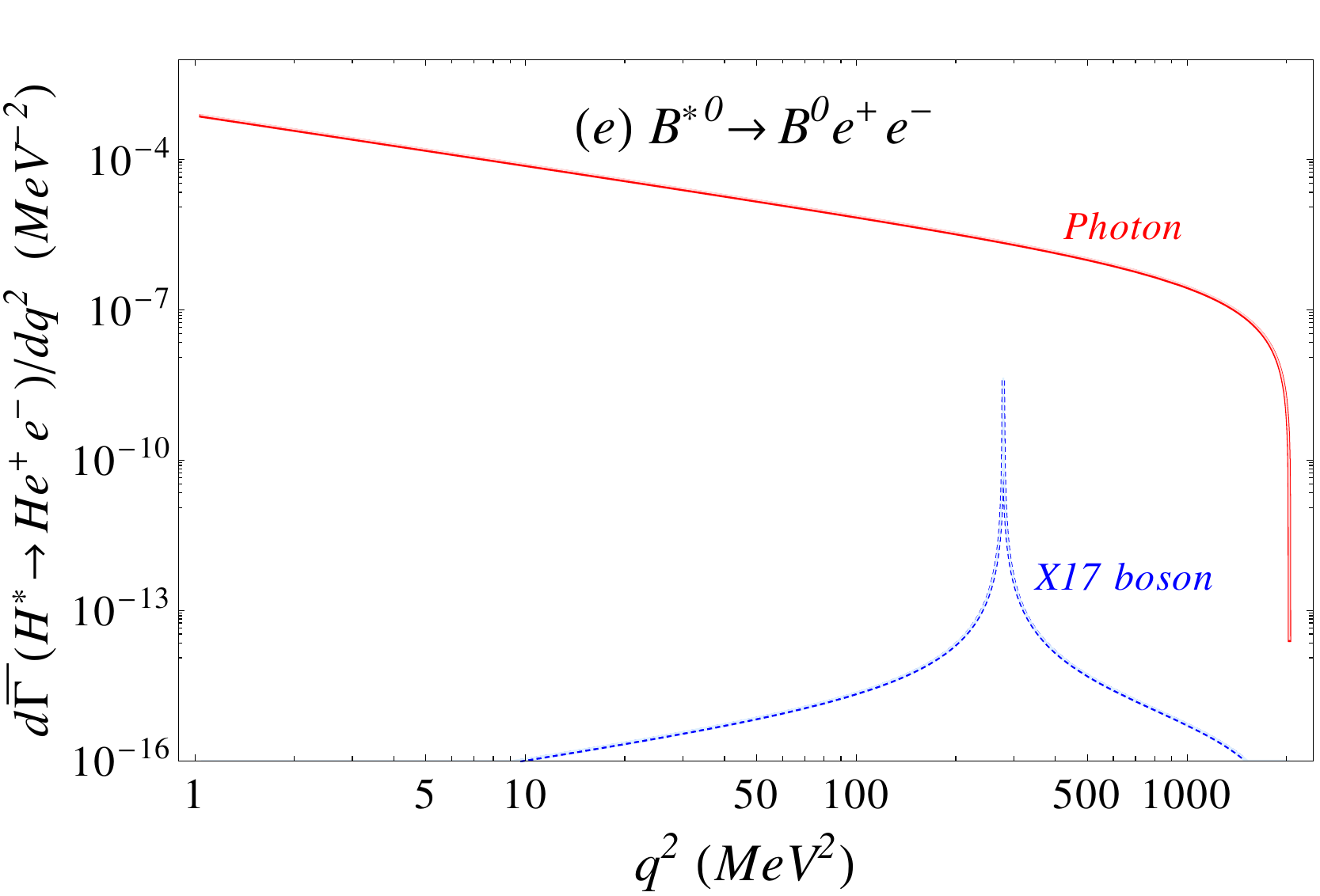} \ \
\includegraphics[scale=0.25]{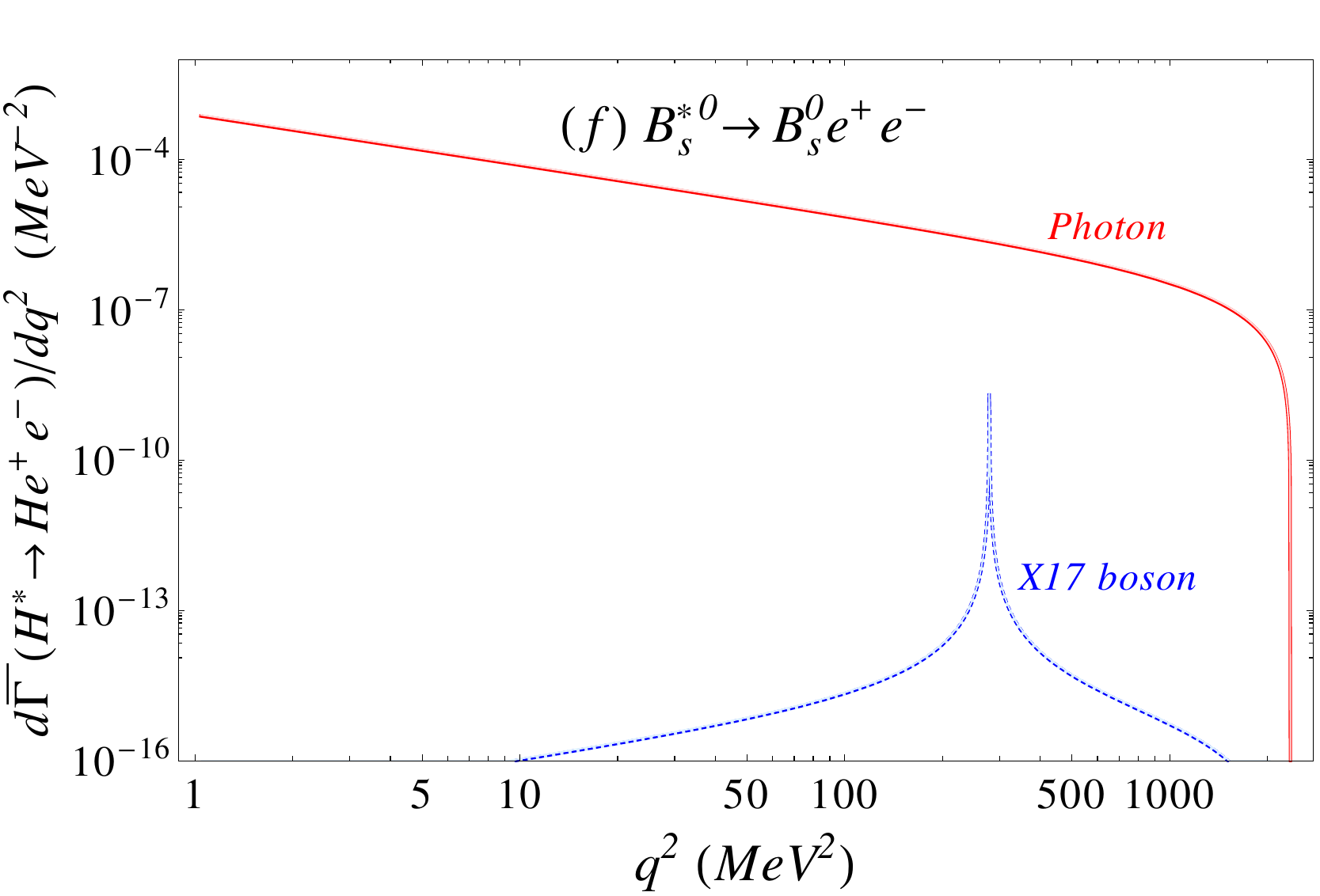}
\caption{\small Lepton pair invariant mass distributions of $H^\ast \to H e^+e^-$ transitions normalized to the radiative $H^* \to H\gamma$ decay width: (a) $D^{\ast +} \to D^{+} e^+e^-$, (b) $D^{\ast 0} \to D^{0} e^+e^-$, (c) $D_s^{\ast +} \to D_s^{+} e^+e^-$, (d) $B^{\ast +} \to B^{+} e^+e^-$, (e) $B^{\ast 0} \to B^{0} e^+e^-$ and (f) $B_s^{\ast 0} \to B_s^{0} e^+e^-$. The red-solid plot denotes the virtual photon contribution, while the $X17$ boson contribution is represented by the blue-dashed curve. The (almost invisible) shaded bands account for the theoretical uncertainties in form factors.}
\label{Fig1}
\end{figure*}

The lepton pair invariant mass distribution, normalized to the radiative decay width of $H^*\to H\gamma$, becomes the sum of the photon and $X$-boson mediated distributions, namely  (we use $\lambda(x, y,z)=x^2+y^2+z^2-2xy-2xz-2yz$):
\begin{eqnarray}
&& \frac{d\Gamma(H^\ast \to H e^+e^-)}{dq^2} = \frac{\alpha_{\rm em}^2}{72\pi \Gamma(H^*\to H\gamma)}  \Big|G_{H^*H\gamma}(q^2)  \nonumber \\ 
&& \ \ \ \ \  + G_{H^*HX}(q^2)\Big|^2  q^2  \bigg[ \frac{\lambda(m_{H^\ast}^2,m_H^2,q^2)^{1/2}}{m_{H^\ast}}  \bigg]^3 . 
 \label{DDW_X}
\end{eqnarray}
Given the very narrow width of the $X17$-boson, the interference of the amplitudes is negligible. Indeed,  the interference in the di-lepton spectrum vanishes at the position of the $X17$ and it is suppressed by more than six orders of magnitude relative to the one-photon contribution outside the resonance.

\begin{table*}[!t]
\centering
\renewcommand{\arraystretch}{1.2}
\renewcommand{\arrayrulewidth}{0.8pt}
\begin{tabular}{ccccc}
\hline\hline
Channel & $R^{\gamma}_{ee}(H^*)$ & $R^{X}_{ee}(H^*)$ & Total  & Experiment \\
\hline
$D^{\ast +} \to D^{+} e^+e^-$ & \ \ $6.67 \times 10^{-3} $ & \ \ $(1.05\pm 0.07) \times 10^{-3}$ & \ \ $(7.72 \pm 0.07) \times 10^{-3}$ & $--$ \\
$D^{\ast 0} \to D^{0} e^+e^-$ & \ \ $6.67 \times 10^{-3} $ & \ \ $3.02 \times 10^{-5}$ & \ \ $6.70  \times 10^{-3}$ & $--$\\
$D_s^{\ast +} \to D_s^{+} e^+e^-$ & \ \ $6.72 \times 10^{-3} $ &  \ \ $(3.10 \pm 0.60) \times 10^{-3}$ & \ \ $(9.82 \pm 0.60) \times 10^{-3}$ & \ \ $(7.2^{+1.8}_{-1.6})\times 10^{-3}$~\cite{CroninHennessy:2011xp} \\
$B^{\ast +} \to B^{+} e^+e^-$ & \ \ $4.88 \times 10^{-3} $ & \ \ $(1.91 \pm 0.03) \times 10^{-5}$ & \ \ $4.90\times 10^{-3}$ & $--$\\
$B^{\ast 0} \to B^{0} e^+e^-$ & \ \ $4.88 \times 10^{-3} $ & \ \ $3.96 \times 10^{-4} $ &  \ \ $5.28 \times 10^{-3}$ & $--$\\
$B_s^{\ast 0} \to B_s^{0} e^+e^-$ & \ \ $4.99 \times 10^{-3} $ & \ \ $4.08\times 10^{-4} $ &  \ \ $5.40 \times 10^{-3}$ & $--$\\
\hline\hline
\end{tabular} 
\caption{\small  Photon and $X17$ boson exchange contributions to the ratio of  decay rates defined in Eq. (\ref{ratioee}). We assume universal couplings of the hypothetical $X17$ boson to down-type quarks [$\varepsilon_b=\varepsilon_s=\varepsilon_d=\mp 7.4\times 10^{-3}$] and up-type quarks [$\varepsilon_c=\varepsilon_u=\pm 3.7\times 10^{-3}$] (see end of Section II). Unless explicityly indicated, theoretical uncertainties are at least three-orders of magnitude smaller than the corresponding central values.}
\label{Table:3}
\end{table*}

The lepton-pair invariant mass distributions due to photon (solid-red) and $X17$-boson (dashed-blue) exchange are shown separately in Figure~\ref{Fig1} for the six different decay channels under consideration. The shaded bands around each curve represents the theoretical error, which are difficult to visualise  in the log-scale. The  peak due to the production of the $X17$ boson in each channel is not located very close to the end of the lepton-pair spectrum as it happens in the nuclear case, avoiding in this way possible end-point kinematical effects. In contradistinction to the on-shell $X17$ production, the effect of this boson is the largest for the $D^{*+}(D_s^{*+}) \to D^+(D_s^+)e^+e^-$ decay.  
The corresponding peaks  of this boson contribution is suppressed by one or two orders of magnitude in all other cases, relative to the photon contribution. Note that we are assuming universality bounds on heavier quark $\varepsilon_{c,s,b}$ couplings; since this is a conservative assumption,  the experimental study of heavy mesons transitions involving lepton pairs may serve to set bounds on these unknown couplings of the hypothetical $X17$ boson.

Table~\ref{Table:3} displays the values of the decay rates for the lepton-pair production in $H^*\to H$ transitions normalized to the corresponding  rates of the radiative decays $H^* \to H\gamma$, namely
\begin{equation}
R_{ee}(H^*)\equiv  \frac{ \Gamma(H^* \to H e^+e^-)}{\Gamma(H^*\to H\gamma)}
\label{ratioee}
\end{equation}
where the radiative rate  is given by $\Gamma(H^* \to H\gamma) = (\alpha_{\rm em}/3) \vert  F_{H^\ast H\gamma}(0)\vert^2 |\vec{p}_\gamma|^3$.  We expect that the remaining model-dependent terms in the form factors are cancelled in this ratio (all other lepton-pair and angular distributions in the following are normalized to this radiative width). As in the case of the lepton-pair spectra, the largest contribution of the $X17$ boson is observed for the $D^{*+}$ and $D_s^{*+}$ decays, making these channels the most sensitive for the observation of this light boson effects. Our calculation of the electromagnetic contribution in the case of $D^*_s$ decays yields $R^{\gamma}_{ee}(D_s^{*+})=6.8\times 10^{-3}$ is in good agreement with the experimental value $(7.2^{\ +1.8}_{\ -1.6})\times 10^{-3}$ reported in~\cite{CroninHennessy:2011xp}. When we add the contribution of the $X17$ boson exchange, our prediction increases to $R^{\gamma+X}_{ee}(D^*_s)= (9.8\pm 0.6)\times 10^{-3}$, which exceeds the experimental value but it is still consistent with it within 1.4$\sigma$.
 Let us notice that a previous prediction of this ratio $R_{ee}(D^*_s)= 6.5 \times 10^{-3}$ was estimated in Ref.~\cite{CroninHennessy:2011xp} based on the model proposed in \cite{Landsberg:1986fd} which includes only the electromagnetic contribution.

\begin{figure}[!t]
\centering 
\includegraphics[scale=0.38]{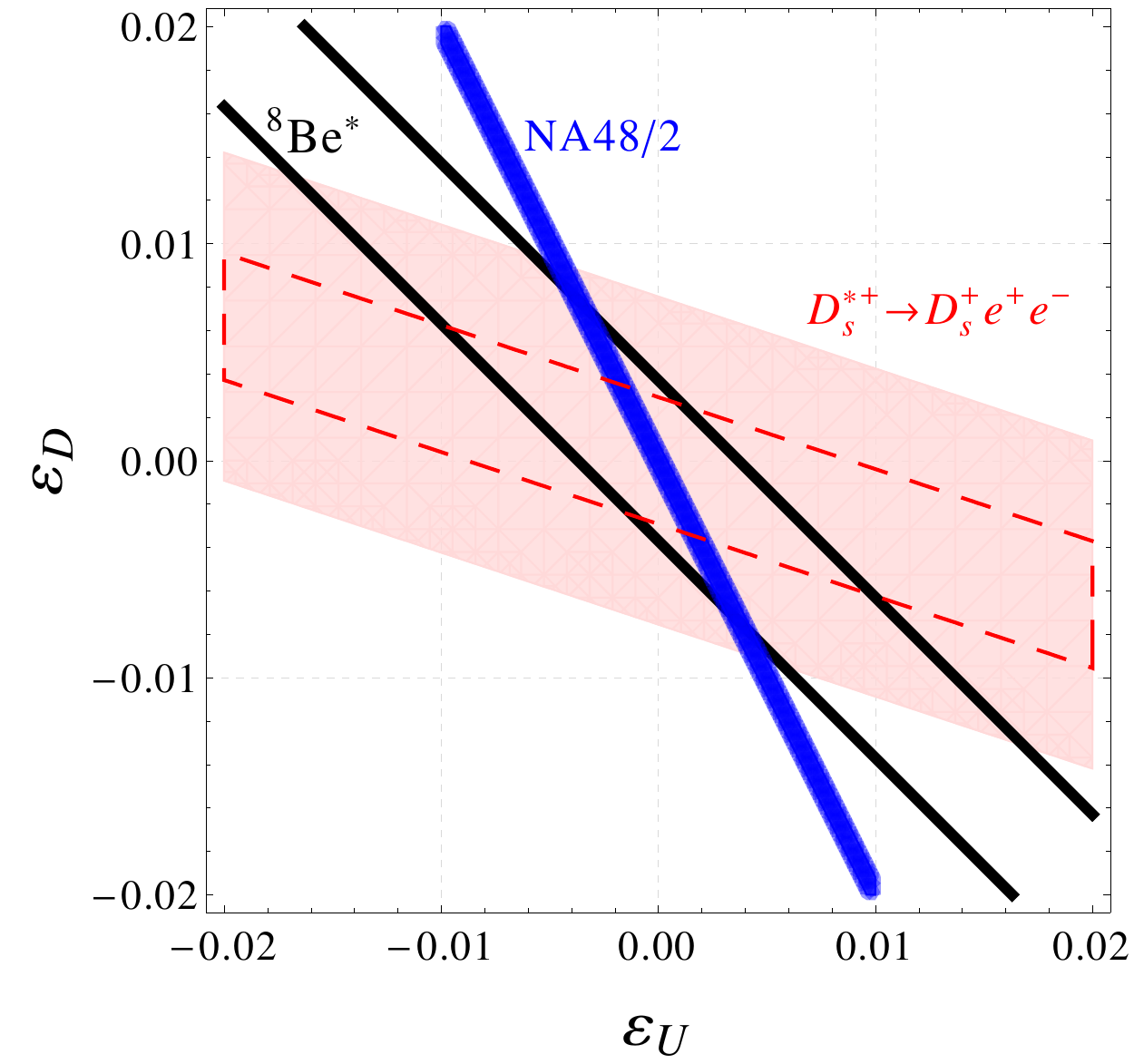}
\caption{\small The $1\sigma$ confidence level allowed regions in the parameter space of up-type ($\varepsilon_U=\varepsilon_c$) and down-type ($\varepsilon_D=\varepsilon_s$) $X$-quark couplings from $D_s^{\ast +} \to D_s^{+} e^+e^-$ (light-red shaded band). The region within dashed lines corresponds to the assumption of a five-fold improvement in the experimental uncertainty. The  corresponding constraints on ($\varepsilon_U=\varepsilon_u, \varepsilon_D=\varepsilon_d$) from the experimental results from $^8$Be$^*$~\cite{Krasznahorkay:2015iga,Feng:2016jff,Feng:2016ysn} and NA48/2~\cite{Feng:2016ysn,Batley:2015lha} are represented by the the two paralell thin black lines and the wider steepest blue band, respectively.}
\label{Fig2}
\end{figure}
  
sThe sensitivity of $D_s^*$ decays into lepton pairs to the effects of $X17$ boson exchange observed in the previous paragraph, suggests this channel can be useful to constrain the parameter space of the hypothetical vector boson. In Fig.~\ref{Fig2} we show the $1\sigma$ confidence level allowed for the parameter space in the $(\varepsilon_c, \varepsilon_s)$ plane, obtained from the comparison of the experimental branching fraction  reported by CLEO~\cite{CroninHennessy:2011xp}
 and the result of integrating Eq. (\ref{DDW_X}) for $D_s^*\to D_se^+e^-$ (light-red shaded band). The current experimental uncertainty in $R(D_s^*)$ is close to $25\%$, and current experiments producing a large dataset of charmed mesons have not  planned new measurements. Therefore, we will assume that a dedicated measurement of this observable may reach an improvement of the current uncertainty by a factor of five. Under this assumption we get the region enclosed by the red-dashed contour in Figure \ref{Fig2}. For comparison, we also show the two thin parallel black lines corresponding to the allowed values of ($\varepsilon_u, \varepsilon_d$) obtained from $^8{\rm Be}^*$ results \cite{Feng:2016jff,Feng:2016ysn}and the region allowed from the so-called `protophobic condition' obtained from the non-observation of $\pi^0\to \gamma X$ by the  NA48/2 experiment  \cite{Feng:2016ysn,Batley:2015lha} (single steepest blue band). The different sensitivities observed from these measurements to the  up-type and down-type quark couplings makes worth an improved measurement of the heavy mesons decays discussed in this paper.

Finally, let us comment that the angular distribution of the $e^+e^-$ pair, in the rest frame of the decaying particle, will be peaked closer to the collinear configuration compared to the nuclear case of $^8$Be$^*$ transitions, where $\theta(e^+e^-)\sim 140^0$. This happens because the $X17$ boson is produced with a larger velocity, while in nuclear transitions this boson is produced almost at rest. 

\section{Conclusions}

The hypothetical light vector boson $X17$, proposed as a solution for the anomaly observed in lepton-pair production of $^8$Be$^*$ and $^4$He transitions, can be studied in the clean environment provided by vector to pseudoscalar  heavy mesons transitions in Belle, Belle II and BESIII factories. These $H^*(Q\bar{q})\to H(Q\bar{q})e^+e^-$ decays are free from theoretical uncertainties associated to  nuclear effects. We have used the HQET+VMD framework to model the hadronic form factors of $1^-\to 0^-$ meson transitions, however our results are little-dependent on hadronic uncertainties because the rates are normalized to the dominant $H^* \to H\gamma$ electromagnetic decays and the dominant contributions in most channels are dominated by photon emission off the light quarks in this model. 

Although all the branching fractions of the  heavy meson channels considered in this paper exhibit some sensitivity to the effects of the $X17$ boson, decays of $D^{*+}$ and $D_s^{*+}$ mesons turn out to be the most sensitive ones. This happens because 1) the radiative charged charmed vector meson decay rates used  as a normalization factor in $R_{ee}(D_s^*)$ and $R_{ee}(D^{*+})$ are suppressed in the HQET+VMD owing to a partial cancellation of the heavy and light quarks contributions and, 2) the large contribution of the light quark coupling to $X17$ for $D^{*+}\to D^+$ transition. Also, improved measurements of these leptonic decay channels can set additional and complementary  constraints on the $X17$ boson couplings to ordinary fermions, as shown in Figure \ref{Fig2} for the case of $D_s^* \to D_se^+e^-$ decays or, eventually, confirm the existence of this light boson.

\acknowledgments
GLC acknowledges support from Ciencia de Frontera project No. 428218 (Conacyt). The work of N. Quintero has been financially supported by MINCIENCIAS and Universidad del Tolima through Convocatoria Estancias Postdoctorales No. 848-2019 (Contract No. 834-2020), and Direcci\'{o}n General de Investigaciones - Universidad Santiago de Cali under Project No. 935-621118-3.





\end{document}